# Transfer learning: Improving neural network based prediction of earthquake ground shaking for an area with insufficient training data

Dario Jozinović[1,2], Anthony Lomax[3], Ivan Štajduhar[4], Alberto Michelini[1]

[1] *Istituto Nazionale di Geofisica e Vulcanologia, Via di Vigna Murata 605, 00143 Rome, Italy*
[2] *Department of Science, Università degli Studi Roma Tre, Via Ostiense, 159, 00154 Rome, Italy*
[3] *ALomax Scientific, 320 Chemin des Indes, 06370 Mouans-Sartoux, France*
[4] *Department of Computer Engineering, Faculty of Engineering, University of Rijeka, 51000 Rijeka, Croatia*

## Summary

In a recent study (Jozinović et al, 2020) we showed that convolutional neural networks (CNNs) applied to network seismic traces can be used for rapid prediction of earthquake peak ground motion intensity measures (IMs) at distant stations using only recordings from stations near the epicenter. The predictions are made without any previous knowledge concerning the earthquake location and magnitude. This approach differs significantly from the standard procedure adopted by earthquake early warning systems (EEWSs) that rely on location and magnitude information. In the previous study, we used 10 s, raw, multistation (39 stations) waveforms for the 2016 earthquake sequence in central Italy for 915 M ≥ 3.0 events (CI dataset). The CI dataset has a large number of spatially concentrated earthquakes and a dense network of stations. In this work, we applied the same CNN model to an area around the VIRGO gravitational waves observatory sited near Pisa, Italy. This large infrastructure can greatly benefit from an EEWS to shut-off the data acquisition in case of significant earthquakes in the vicinity. In our initial application of the technique, we used a dataset consisting of 266 M ≥ 3.0 earthquakes recorded by 39 stations. We found that the CNN model trained using this smaller-sized dataset performed worse compared to the results presented in the original study by Jozinović et al. (2020). To counter the lack of data, we explored the adoption of "transfer learning" (TL) methodologies using two approaches: first, by using a pre-trained model built on the CI dataset and, next, by using a pre-trained model built on a different (seismological) problem that has a larger dataset available for training. We show that the use of TL improves the results in terms of outliers, bias and variability of the residuals between predicted and true IMs values. We also demonstrate that adding knowledge of station positions as an additional layer in the neural network improves the results. The possible use for EEW is demonstrated by the times for the warnings that would be received at the station PII by using the CNN model.

# 1 Introduction

Having information about earthquake generated ground motions in the shortest time possible (Minson et al., 2018) is of great importance for earthquake monitoring. For a time scale of 5-10 minutes, after the earthquake, the ShakeMap software (Wald et al., 1999) was developed to provide maps of ground shaking (in terms of peak ground acceleration (PGA), peak ground velocity (PGV), spectral accelerations at 0.3s, 1s and 3s periods and macroseismic intensity), which are then used by disaster risk managers for assessments of shaking impact in the area and further action based on the assessment. On a shorter time scale (few seconds) earthquake early warning systems (EEWS) have been developed in the seismological community (see the reviews of Satriano et al. 2011, Cremen et al. 2020). These systems, developed as regional (e.g. Kohler et al. 2017) or on-site EEWS (e.g. Spallarossa et al. 2019), seek to detect and characterize earthquakes rapidly and provide warnings to points, or areas, not yet impacted by ground shaking.

In a recent study, Jozinović et al. (2020) (J2020, hereinafter) used a machine learning (ML) approach to predict the ground shaking intensity at a predefined set of seismic stations within a given seismic area, as quickly as possible. The inputs to the ML model were 7 s, 10 s, or 15 s long waveforms (the length of the waveform window was investigated within the study) from a predefined and fixed set of seismic stations with all traces starting at the earthquake origin time for simplicity. The outputs of the ML model were the intensity measures (IMs; i.e., PGA, PGV and SA at 0.3 s, 1 s, 3 s periods) on the selected stations. This configuration entails that the strongest shaking would be recorded on the traces of the stations closest to the epicenter, while the model would give predictions for the stations farther from the epicenter. Clearly, the balance between the recorded and predicted IMs will depend upon the source-receiver acquisition geometry and the selected traces' window length. One notable feature of the model is that it does not require any information about the earthquake parameters (magnitude, location, etc.), and uses just the multistation waveforms pattern to give the predictions. The best compromise between the accuracy and timeliness was found when 10 s windows were used. In J2020 it was also found that the IMs prediction accuracy was similar to the accuracy attainable using the ground motion prediction equations (GMPEs) by Bindi et al. (2011) which require, however, an earthquake location and magnitude as input. In addition, it was found that the ML model was able to predict with useful accuracy the IMs at the stations which had no input data available (and were replaced with a window of zeros).

In this study, we use the model developed in J2020 and apply it to the area of central-western Italy (Fig. 2), centered at the VIRGO gravitational wave observatory near Pisa. The VIRGO detector site would benefit from an EEWS (Olivieri et al., 2019) with prediction of IMs to prevent damage on the instrument or the shut-off of the data acquisition (to avoid losing the resonant condition of the instrument) for significant earthquakes nearby. We use VIRGO as an example of an infrastructure that could benefit from the adoption of our ML-based methodology in a EEWS fashion.

Increasing use of ML in seismology has led to the development of ML based rapid earthquake characterization algorithms and rapid peak ground motion prediction algorithms. Some of the developed algorithms deal with rapid seismic wave discrimination (e.g. Li et al., 2018) or rapid earthquake characterization (e.g. Böse et al. 2012, Hsu et al. 2016, Ochoa et al. 2018, Saad et al. 2020, van den Ende and Ampuero 2020, Münchmeyer et al. 2021, Zhang et al. 2021). Böse et al. (2008) have approached the problem of rapid earthquake characterization for EEWS using multi-station waveforms to extract a series of chosen parameters and use them as inputs for a feedforward neural network. Kong et al. (2016)

developed a ML algorithm that detects earthquakes on smartphone accelerometers and uses the information from the triggered smartphones to estimate the earthquake location. Otake et al. (2020) used a recurrent neural network that adopted waveforms from 4 stations as input, to predict the shaking intensity at one target location. The study of Münchmeyer et al. (2020) deals with the EEWS problem with a technique called TEAM, which takes multi-station seismic waveforms as input and predicts the PGA, which is similar to the technique developed in J2020 and used in this paper. The main difference between the two algorithms derives from the use of different neural network model types. TEAM uses a combination of a transformer and a convolutional neural network (CNN) whereas in J2020 we rely exclusively on a CNN. The advantage of using TEAM comes from the possibility to use any set of up to 25 stations as the input to the model, which provides flexibility in applying the network to different areas without the need to retrain the model. In contrast, the J2020 approach always uses the same, structured, set of stations as the input to the model. The advantage of this approach is that the CNN learns the specifics of ground motion at every station and the whole pattern amongst the stations but it needs to be retrained in order to be applied to a different area with a different set of stations. This would not be a limit for the J2020 model for application to a dense set of stations, like those in J2020, with a large enough training dataset. However, training it for other areas with a sparse network of stations that has a smaller-sized dataset of earthquake waveforms available for training, would likely lead to poorer results compared to J2020. In this study, we try to overcome this problem by testing the use of Transfer Learning (TL) for model training (Bozinovski, 2020; Pan et al., 2010 for a review).

TL uses two ML models in the following way: the first, already trained, model is used for initialization of the weights of the second model. More precisely, TL consists of taking a model pre-trained on a source dataset (usually larger) or a source task (e.g. single-station magnitude determination) and using it (or its parts) for training a model on a target dataset or for a target task (e.g. multi-station IM prediction), where the source and target datasets, or tasks, are sampled from different underlying distributions. In doing this, it is expected that the first distribution is similar enough to the second, most importantly in the input datasets, so that its use for TL will likely improve model performance compared to a model having its weights randomly initialized. Note that TL is effectively the replication of the natural mental processes that occur in all species of exploiting previous knowledge for new learning.

There has been some application of TL to seismology (e.g., Chai et al. 2020, Titos et al. 2019, Johnson et al. 2021, Otović et al. 2021) although its use is not yet widespread for applications of ML in seismology. In this study, we explore the use of TL for improving our IM prediction algorithm. We use the pre-trained model from the J2020 study, with the same source task as the task used in this study (which means that we use the same model architecture as the one in J2020), but with a different source dataset. We also explore TL from a different source task (magnitude determination from single station waveforms), trained on a different source dataset.

## 2 Data

The input data from central western Italy (hereinafter denominated the CW data set) consist of 3-component waveforms of 256 earthquakes with magnitude $2.9 \leq M \leq 5.1$ (Fig. 1a), from 39 stations. The earthquakes occurred between 1 January 2013 and 20 November 2017. The earthquake depths range from 3.3 to 64.7 km (Fig. 1b). The stations and the earthquakes are located in the area bounded by latitude [41.13°, 46.13°] and longitude [8.5°, 13.1°] (Fig. 2), with epicentral distances ranging from 10 - 498 km (Fig. 1c). The area overlaps slightly with the area of central Italy used in J2020 from which the pre-trained model architecture is taken. To avoid possible data leakage from the pre-trained model, the events that were used in that study were excluded from our dataset. We chose the stations having the largest number of events recorded while making sure that there is an acceptable spatial distribution of the stations to cover more area with stations that are close to earthquake epicenters, to simulate possible early warning use (more details about the stations in Table 1. of the electronic supplement). The stations belong to the IV, GU and MN networks.

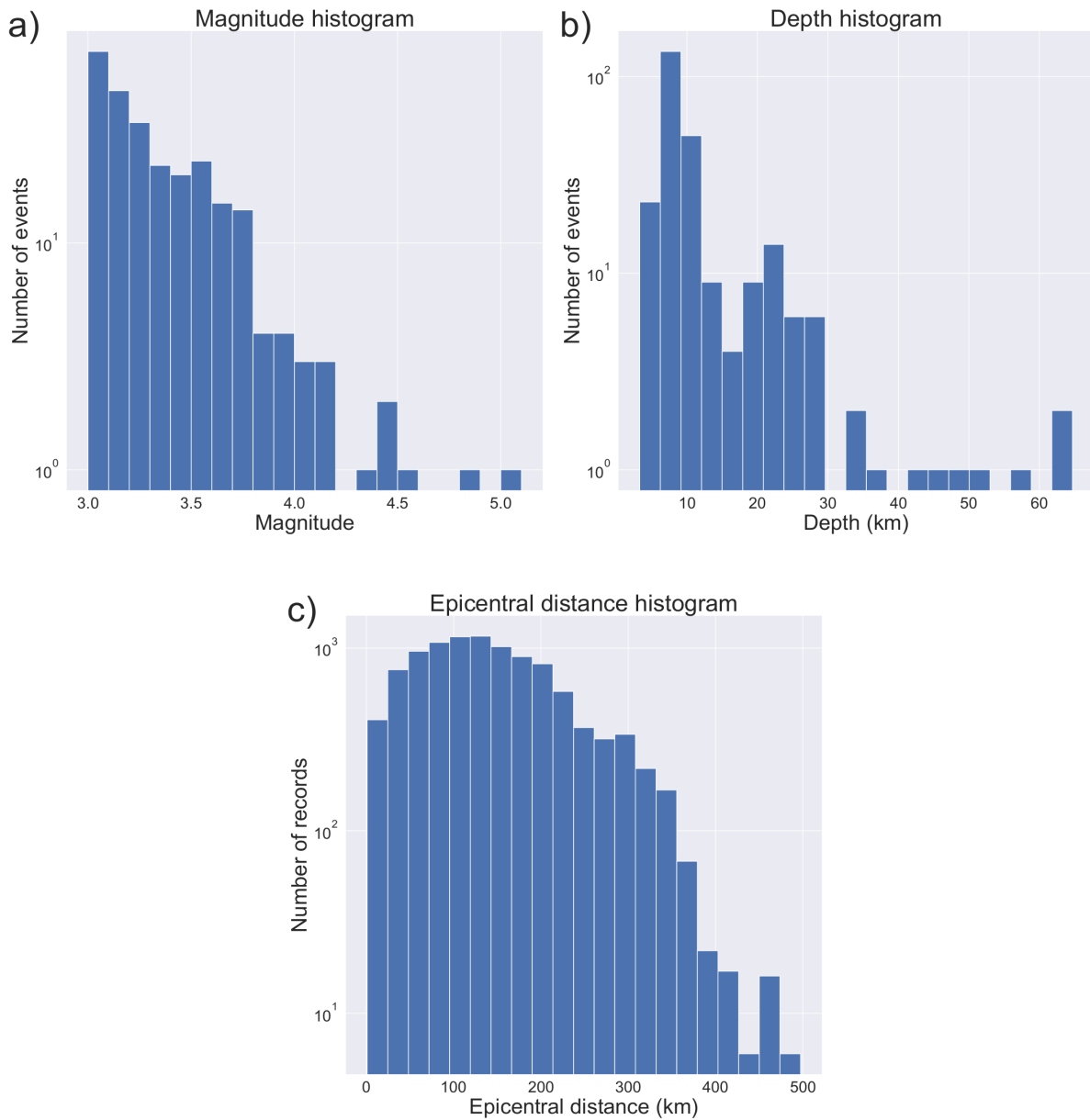

*Figure 1. Histograms of the selected 256 earthquakes: a) Earthquake-Magnitude distribution, b) Earthquake-Depth distribution, c) Epicentral distances distribution*

The three component, 3C, waveform data were downloaded using the INGV FDSN web services for HN* (acceleration) and HH* and EH* (velocity) channels, where * ∈ [E, N, Z]. The data were processed to remove the instrument response, velocity data were differentiated to acceleration and, if necessary, the data were resampled to 100 Hz. For M < 4 earthquakes, the HH and EH channels were used after differentiation and for earthquakes with M ≥ 4.0 the HN channels were used. For certain stations and for some earthquakes, the waveform data were completely missing and we chose to replace them with zeros, as in J2020. This data selection and processing follows the criteria outlined in J2020.

The *target variables* consisted of the IMs associated with each recording: peak ground acceleration (PGA), peak ground velocity (PGV), spectral acceleration (SA) at 0.3 s, 1 s and 3 s periods. For stations that had no data, the IMs were calculated using the USGS ShakeMap software using the latest configuration for Italy (Michelini et al., 2020) to ensure no missing output data (target variables). This resulted in the following composition of target values: 91.4% were observed, while 8.6% were calculated using ShakeMap. We have calculated the first P arrival times on the stations from the theoretical travel times using the Java TauP Toolkit by Crotwell et al. (1999) that calculates theoretical travel times and paths as implemented in the Obspy Python library (Krischer et al., 2015) and the *ak135* velocity model (Kennett et al., 1995). We performed a visual check of several waveforms and it was found that the theoretical travel times calculated were satisfactory to the purpose of this study.

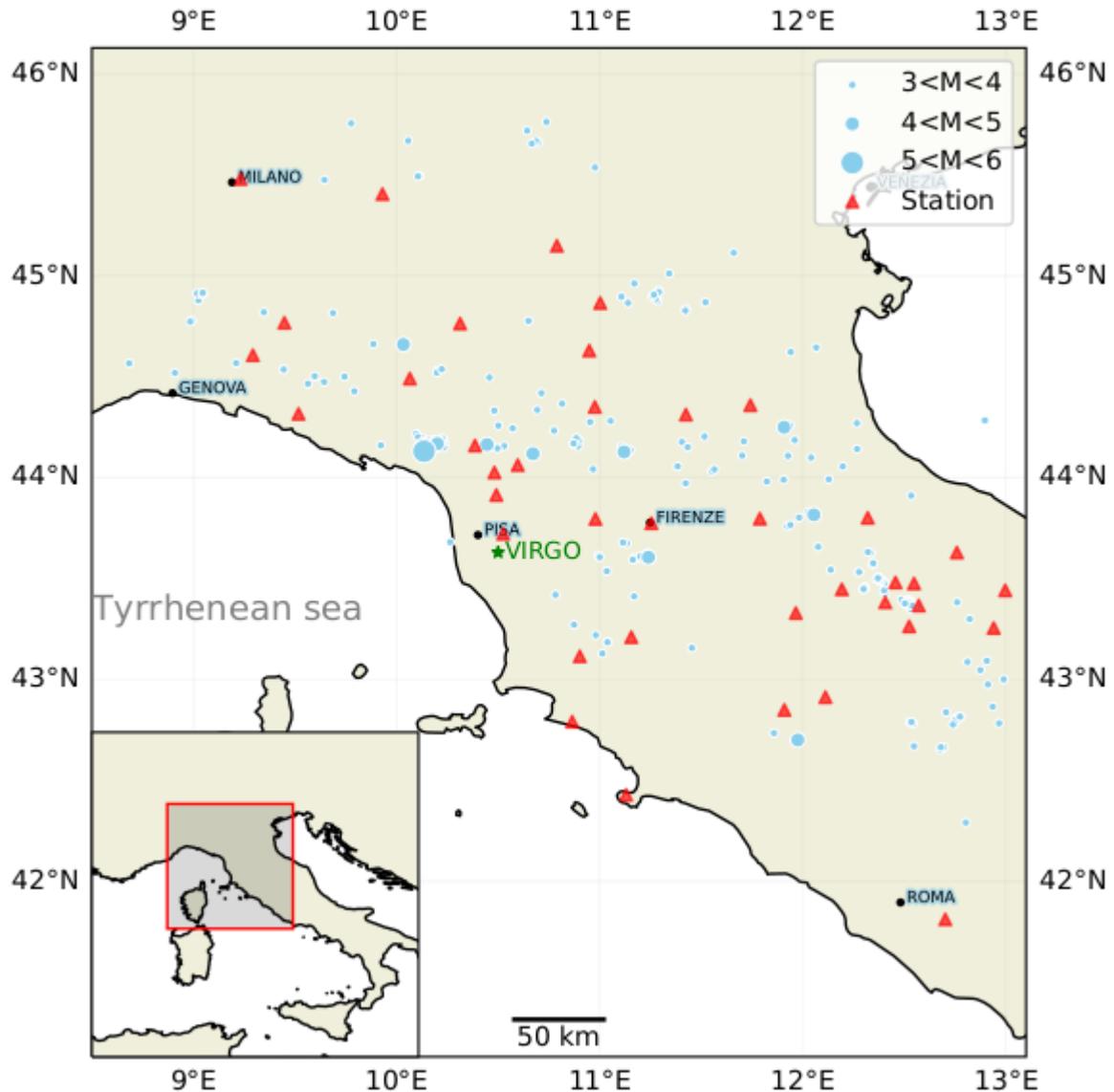

*Figure 2. Spatial distribution of the 256 earthquakes (blue dots - scaled according to earthquake magnitude) together with the 39 stations (red triangles) selected for this study and the location of the VIRGO observatory (green star).*

We use 3 other datasets for pre-training the models for TL. The first dataset is the J2020 central Italy dataset (CI hereinafter) consisting of 915 earthquakes recorded by 39 stations with the same sampling rate and target labels as provided in this study. CI has essentially the same structure as the dataset from this study except for a different set of recording stations.

The other two datasets used are global ones consisting of local earthquakes, LEN-DB (Magrini et al. 2020) and STEAD (Mousavi et al. 2020). These datasets are significantly larger than CI, and they provide 3C single station earthquake waveforms, i.e. the dataset structure is different from the dataset of this study. In particular, the waveforms provided by these datasets are not recordings from a fixed set of stations. This essentially means that we cannot use these datasets to pre-train a CNN model that has the same, multi-station, architecture for IM prediction that we use in this study although they can be used in the first layers of our CNN model as explained below.

The differences in these two large datasets come from the sampling rates used, 20 Hz (LEN-DB) and 100 Hz (STEAD), and the amplitude units of the waveforms, velocity (LEN-DB) and data counts (STEAD). The LEN-DB data were differentiated to acceleration. Another difference between the datasets is the maximum epicentral distance, where the STEAD maximum is 350 km, and LEN-DB maximum is 120 km. In both databases, we used only the earthquake waveforms with magnitude M >= 3 (the criterion adopted also when preparing the dataset for this study) providing 77,487 data from LEN-DB and 106,245 data from STEAD.

## 3 Method and training

The CNN model adopts the architecture proposed by J2020, using the Keras Python library (Chollet et al., 2015). Input to the model is a combination of all the waveform data (all 39 stations) for a given earthquake, for an input array size (39, 1000, 3), where 39 is the number of stations, 1,000 is the number of samples (with a sampling rate of 100 samples per second and a 10 s window) and 3 is the number of components. The ordering of the stations is always preserved. The waveform data for each earthquake starts 3 seconds before the estimated P arrival time at the station nearest to the epicenter or at earthquake origin time if the estimated P arrival is less than 3 seconds after origin time (for consistency with J2020 where all the arrivals at the closest station were 3 seconds after origin time). The data are normalised by the input maximum (i.e., the largest amplitude observed across all stations within the time window), and this maximum is saved as the normalisation value which is later inserted into the network. The model outputs are arrays of size (39,5), where 39 is the number of stations, and 5 is the number of predicted IMs per station (i.e., PGA, PGV, SA at 0.3 s, 1 s and 3 s periods). We applied the base-10 logarithm to all the IMs (i.e., $\log_{10}$IM). Relative to J2020, we made a small change to the architecture - we moved the dropout to the flattened layer, instead of the layer that was combining the metadata and the flattened layer (Fig. 3a). We did this because in one of the tests we add additional 117 (39x3) constant data (the station distances and azimuths to a reference station and the $v_{S,30}$ at the stations) to the metadata layer, and in this way, we ensure that the constant metadata inserted are always present, throughout the model-training process.

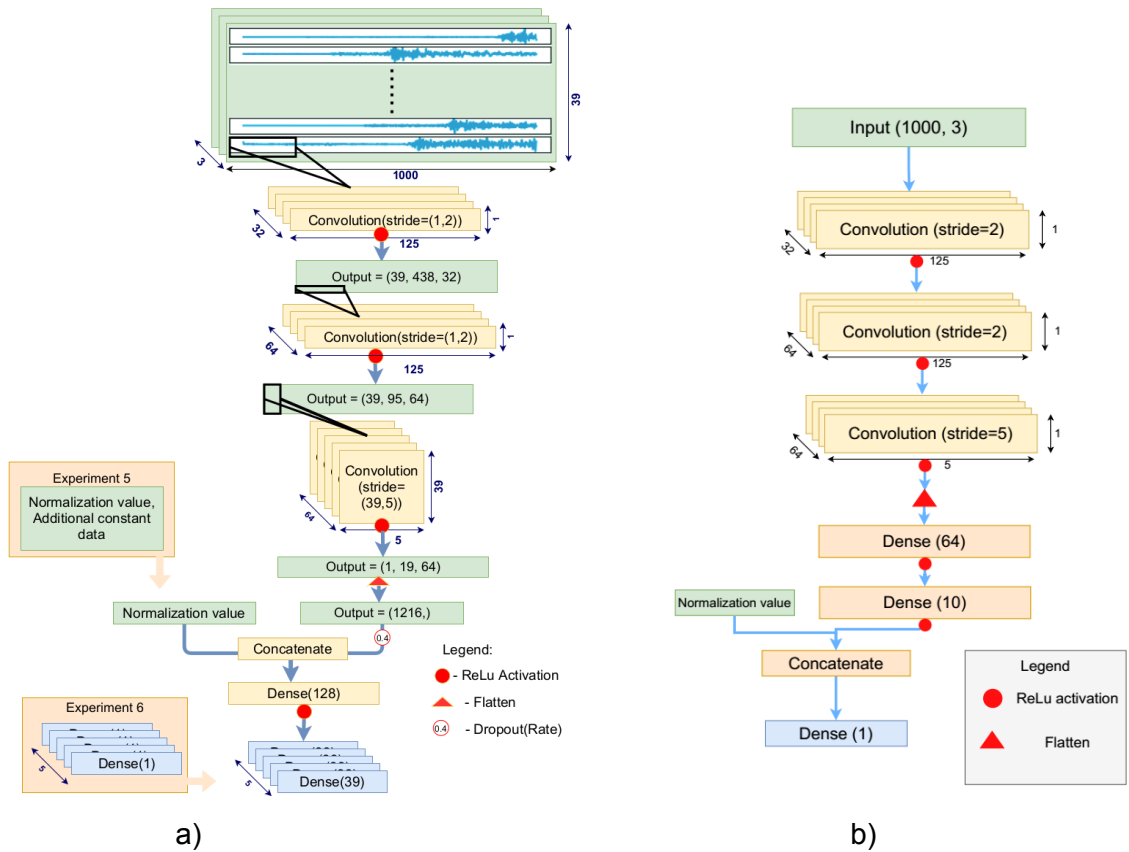

Figure 3. *The architecture of the CNN models used. **a)** Model for IM prediction. Boxes shaded yellow represent filter banks and operators, boxes shaded green represent inputs and the boxes shaded blue represents the final output. Orange boxes represent modifications of the model during the experimentation stage, with the title of the boxes denoting the experiment at hand. **b)** Model for magnitude determination. Boxes shaded green represent inputs, shaded yellow represent convolutional layers, shaded orange represent fully-connected layers and box shaded in blue represents the output.*

To demonstrate that TL is beneficial for model performance when the training data is insufficient, we perform a series of experiments. They are illustrated in Fig. 4 and are explained in more detail below. First, we train the model from the very beginning using only the available CW data (1st experiment). Next, we experiment with TL by using a model pre-trained on the same type of problem (CI dataset, 2nd experiment), and with TL from a pre-trained model on a different seismological problem performed in two steps (magnitude characterization; 3rd and 4th experiment). After that, we test the addition of station ($v_{S,30}$) and inter-station information (distances and azimuths) as additional inputs to the CNN model (5th experiment). In the last test, we change the output of the CNN to predict the IMs only at one station (6th experiment). Specific modelling choices were made by trial-and-error while observing model performance on the validation subset.

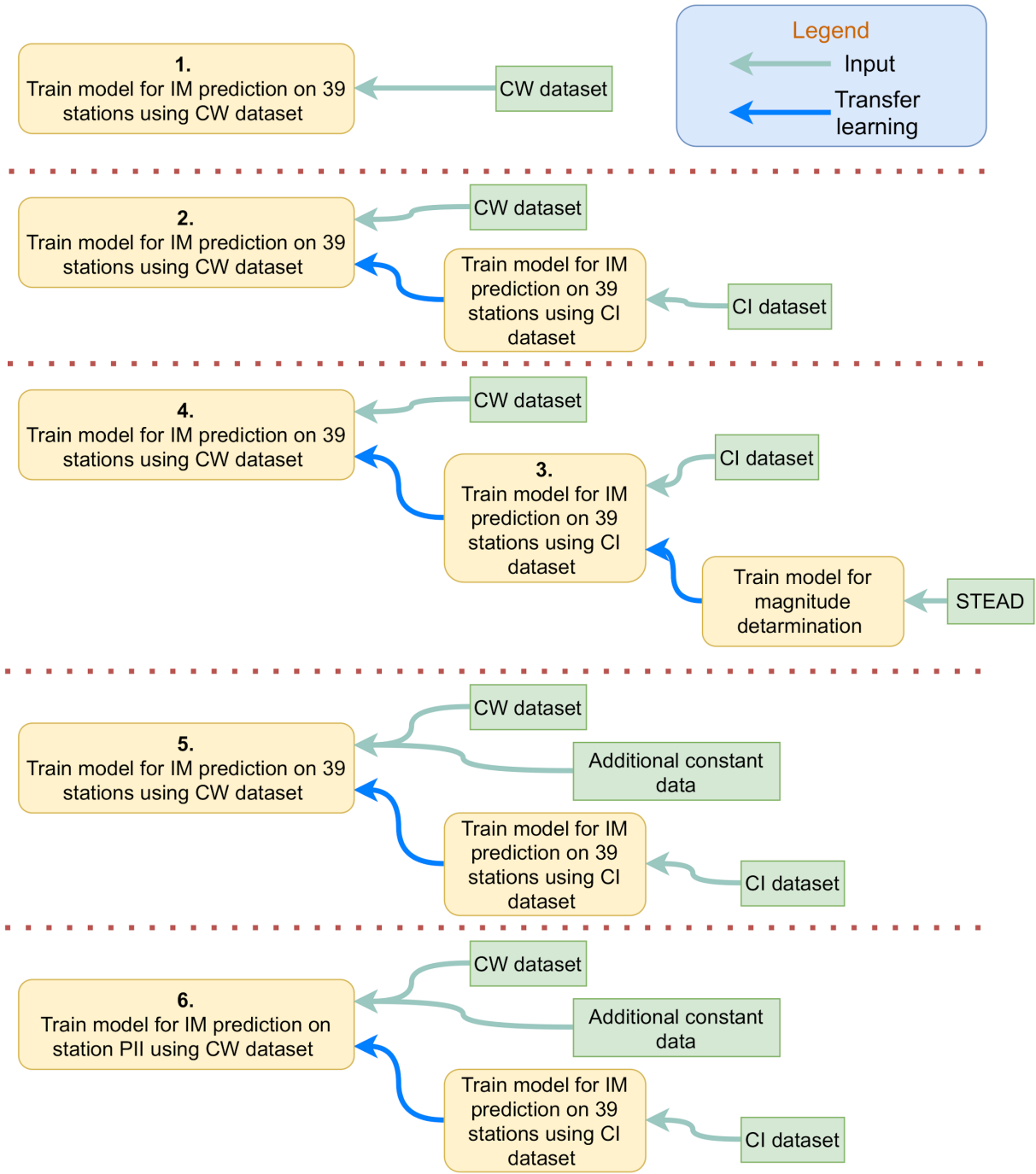

Figure 4. *Diagram of the tests performed. The yellow boxes denote the trained CNN models, with the number in the box marking the number of the test. Input data are represented by the green boxes. Arrows represent input (green) and TL (blue).*

To train the CNN with random weights initialization (1$^{st}$ experiment), the weights are initialized using the *Glorot* uniform initializer (Glorot et al. 2010).

For TL from a pre-trained model, the initial model weights are taken directly from parts of the CNN model from J2020 (2$^{nd}$ experiment). More specifically, only the weights of the first two convolutional layers are used, while the remaining layers are initialized using the Glorot uniform initializer. This has been done since the first two layers in this architecture extract seismogram features irrespective of the station specifics (geographical pattern, soil, topography, etc.), while the subsequent layers combine the extracted features from individual stations exploiting the network station pattern.

The dataset used in J2020 is relatively small, having only 915 earthquakes. Therefore, we use the two other large databases of earthquake waveforms, LEN-DB and STEAD, which provide a large number of training data to improve the feature extractors (i.e., the first two convolutional layers) and exploit TL. To this end, we constructed a CNN model for magnitude determination from single-station earthquake waveforms (Fig. 3b), designing an architecture in which the filters from the first two layers are easily transferable for our task. Then we used the models pre-trained on LEN-DB and STEAD to train the model for central Italy IM prediction (3$^{rd}$ experiment, i.e., improving the results achieved in the J2020 study). When using the LEN-DB data, which have a sampling rate of 20 Hz, for pre-training, the CI data were resampled to 20 Hz and the architecture was adapted to this change. After the validation loss stops improving, all the layers are then fine-tuned, i.e., set as trainable with a small learning rate of 0.00001, and the training is continued. The CI CNN model trained using TL was then used (i.e., TL applied again) as a pre-trained model for training with the CW dataset (4th experiment).

In the next stage of experimentation (5th experiment) we added the information about the stations (interstation distances and azimuths, $v_{S,30}$ of the stations) to the metadata layer (Fig. 3a). The distances and azimuths of every station from an arbitrarily chosen reference station (ASQU) were provided.

In the last part of the experimentation we use only a single station for output (Fig. 3a), station PII, the closest to the VIRGO observatory, to see how much our algorithm could be improved when the target task was "simplified" (6th experiment). Separately, we also experimented with the adoption of shorter time windows.

The data of the CW dataset had been split randomly into training (80%) and test (20%) subsets. For evaluation of the performance of the CNN model, we use 5-fold cross-validation, which splits the randomly permuted dataset into 5 equally-sized disjoint subsets, and uses each of them as the test set to the model trained on the remaining 4 subsets. We have done all the modifications in the model on only one subset, and, when finally defined, we run the training on all the subsets. We then join the predictions on the test subsets from all 5 folds. The analysis of the results is presented in the following section.

The batch size used for optimization was 8, and the mean squared error (MSE) loss function was used for model optimization. Hyperparameter values for model optimization were based on J2020. We have also used an early stopping procedure, which terminates the training if the validation loss did not improve for 25 epochs. Together with the early stopping, we also reduce the learning rate by a factor of 0.2 if the validation loss did not improve for 8 epochs, with the minimum possible learning rate being 0.00001. We used an Nvidia 1050 4GB graphics card for training, which took 13 and 5 minutes on average, for training without TL and when TL approach was used, respectively.

# 4 Results

Here, we present the results of all the 6 experiments. For each experiment, the CNN model results from all 5-fold cross-validation test sets are averaged. Residuals R = $\log_{10}(IM_{obs}/IM_{predicted})$ are calculated and the outlier records with residuals (of any IM) |R|>1 are removed. We removed the outliers to quantify the number of seriously erroneous predictions, which have here been defined as those predictions that are at least an order of magnitude (i.e. |R| > 1) different from the observed values. Finally, we calculated the mean, median, and standard deviation of the residuals, having the outliers removed. All the experimental setups are illustrated in Fig. 4. In Fig. 5, we present the results using box-plots for each IM. For experiments 1-5, we report the results of residuals on all the data here, and the results are split into residuals on observed target labels and ShakeMap-derived target labels in the electronic supplement.

## 4.1. Training with random weights initialization

The results of the CNN model trained with random weights initialization (using the Glorot uniform initializer) are presented here. Removal of outlier residuals resulted in keeping 84.32% of the data. The results are shown in Table 1 and Fig. 5. The results separated for the data with observed targets and with ShakeMap calculated targets are available in the electronic supplement section 3.1.

Table 1: IMs' residual R statistics for the 1st experiment: the CNN predictions on the CW dataset with random weights initialization.

| IM | Median | Mean | STD |
| PGA | -0.084 | -0.07 | 0.413 |
| PGV | -0.104 | -0.088 | 0.408 |
| SA(0.3) | -0.073 | -0.058 | 0.411 |
| SA(1.0) | -0.08 | -0.059 | 0.413 |
| SA(3.0) | -0.095 | 0.079 | 0.423 |

## 4.2. Transfer learning from a model pre-trained on central Italy dataset

The results of the CNN model trained using TL from the pre-trained CI CNN model show that removal of the outlier residuals resulted in keeping 87.96% of the data. The results are shown in Table 2. and Fig. 5. The results separated for the data with observed targets and with ShakeMap calculated targets are available in the electronic supplement section 3.2.

When the results are compared with those of the previous section, we can see that improvement comes in the reduction of the number of outliers and a decrease in standard deviation of the residuals, for both the stations having observed data and those having no observed data (i.e., where ShakeMap predictions were used). We achieved the best results when the weights of the first two convolutional layers are used from the pre-trained model with their learning rates set to zero, while the remaining layers are initialized using the Glorot uniform initializer. We have also tried to train the CNN model using TL while leaving all layers to be fully trainable; however, this deteriorated the performance, suggesting that the feature extractors learned in the first two layers of the pre-trained CNN model are useful for the training on this dataset. This means that the first two layers in our architecture, i.e., those that analyze single station waveforms are transferable between models used for training on different areas, but should be trained on large, high-quality datasets.

Table 2: IMs' residual R statistics for the 2nd experiment: the CNN predictions on the CW dataset using TL from the pre-trained CI CNN model.

| IM | Median | Mean | STD |
|---|---|---|---|
| PGA | -0.069 | -0.055 | 0.394 |
| PGV | -0.085 | -0.069 | 0.38 |
| SA(0.3) | -0.064 | -0.044 | 0.401 |
| SA(1.0) | -0.062 | -0.039 | 0.395 |
| SA(3.0) | -0.075 | -0.061 | 0.392 |

## 4.3. Improving the CI model by TL from a different task

The results for CI using a pre-trained model for magnitude determination from STEAD data are presented here. Removal of the outliers resulted in 96.3% data kept. Results are shown in Table 3. Compared to the previous study J2020 (Table 5 of the Electronic supplement), we achieved an improvement in reducing the number of outliers, the standard deviation of the residuals and their bias (i.e., the difference of |R| from 0). The results separated for the data with observed targets and with ShakeMap calculated targets are available in the electronic supplement section 3.3. We also found that the results are slightly better when using the STEAD pre-trained model.

The best results were achieved when the first layer learning rate was set to 0 and the weights of the second layer were initialized with the pre-trained weights but left trainable. We achieved further improvement in the results by fine-tuning the weights of all the layers with a learning rate of 0.00001.

Table 3: IMs' residual R statistics for the 3rd experiment: the CNN predictions on the CI dataset using TL from the pre-trained magnitude determination model.

| IM | Median | Mean | STD |
|---|---|---|---|
| PGA | -0.011 | -0.006 | 0.28 |
| PGV | -0.019 | -0.008 | 0.257 |
| SA(0.3) | -0.005 | -0.003 | 0.284 |
| SA(1.0) | -0.009 | -0.003 | 0.271 |
| SA(3.0) | -0.017 | -0.005 | 0.292 |

## 4.4. Using the newly trained CI model for TL

Since the results obtained for the CI experiment of J2020 have been improved compared to the previous study (section 4.3), we tried to use the improved pre-trained model trained on CI data to improve the TL for the CW dataset. When the outlier residuals were removed 87.44% of the original data were kept. The results are shown in Table 4. and Fig. 5. The results separated for the data with observed targets and with ShakeMap calculated targets are available in the electronic supplement section 3.4. When comparing the results with the results in section 4.2 (pre-trained CI model trained with random weights initialization) we can see that the results were comparable overall. We found a slight improvement in the residual median, mean and standard deviation, but the number of outliers was slightly bigger in the 4th experiment.

Table 4: IMs' residual R statistics for the 4th experiment: the CNN predictions on the CW dataset using TL from the pre-trained CI CNN model.

| IM | Median | Mean | STD |
|---|---|---|---|
| *PGA* | -0.054 | -0.046 | 0.393 |
| *PGV* | -0.066 | -0.056 | 0.38 |
| *SA(0.3)* | -0.041 | -0.033 | 0.398 |
| *SA(1.0)* | -0.05 | -0.035 | 0.39 |
| *SA(3.0)* | -0.056 | -0.045 | 0.385 |

## 4.5. Adding additional knowledge

The results with inter-station distances and azimuths as additional metadata added to the model are reported here. When the outliers were removed 89.45% of data was kept. The results are shown in Table 5 and Fig. 5. The results separated for the data with observed targets and with ShakeMap calculated targets are available in the electronic supplement section 3.5.

Table 5: IMs' residual R statistics for the 5th experiment: the CNN predictions on the CW dataset using TL from the pre-trained CI CNN model, with added interstation distances and azimuths.

| IM | Median | Mean | STD |
|---|---|---|---|
| *PGA* | -0.049 | -0.036 | 0.375 |
| *PGV* | -0.059 | -0.05 | 0.364 |
| *SA(0.3)* | -0.044 | -0.025 | 0.381 |
| *SA(1.0)* | -0.046 | -0.029 | 0.385 |
| *SA(3.0)* | -0.059 | -0.043 | 0.386 |

When comparing with the results of the previous sections on training the model on the CW dataset (sections 4.1, 4.2, 4.4) we see an improvement in the smaller number of outliers and reduced standard deviation.

The best results were achieved after scaling the distance (expressed in km) by dividing by the maximum distance (246 km), and with azimuths specified as sin and cos of the azimuths. The addition of $v_{S,30}$ did not improve the results.

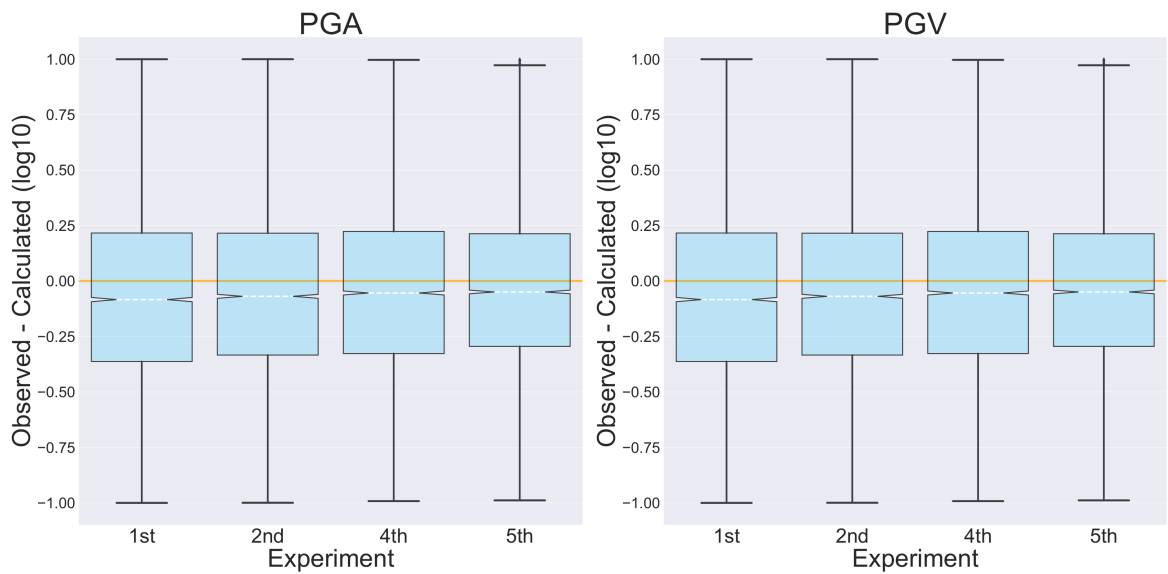

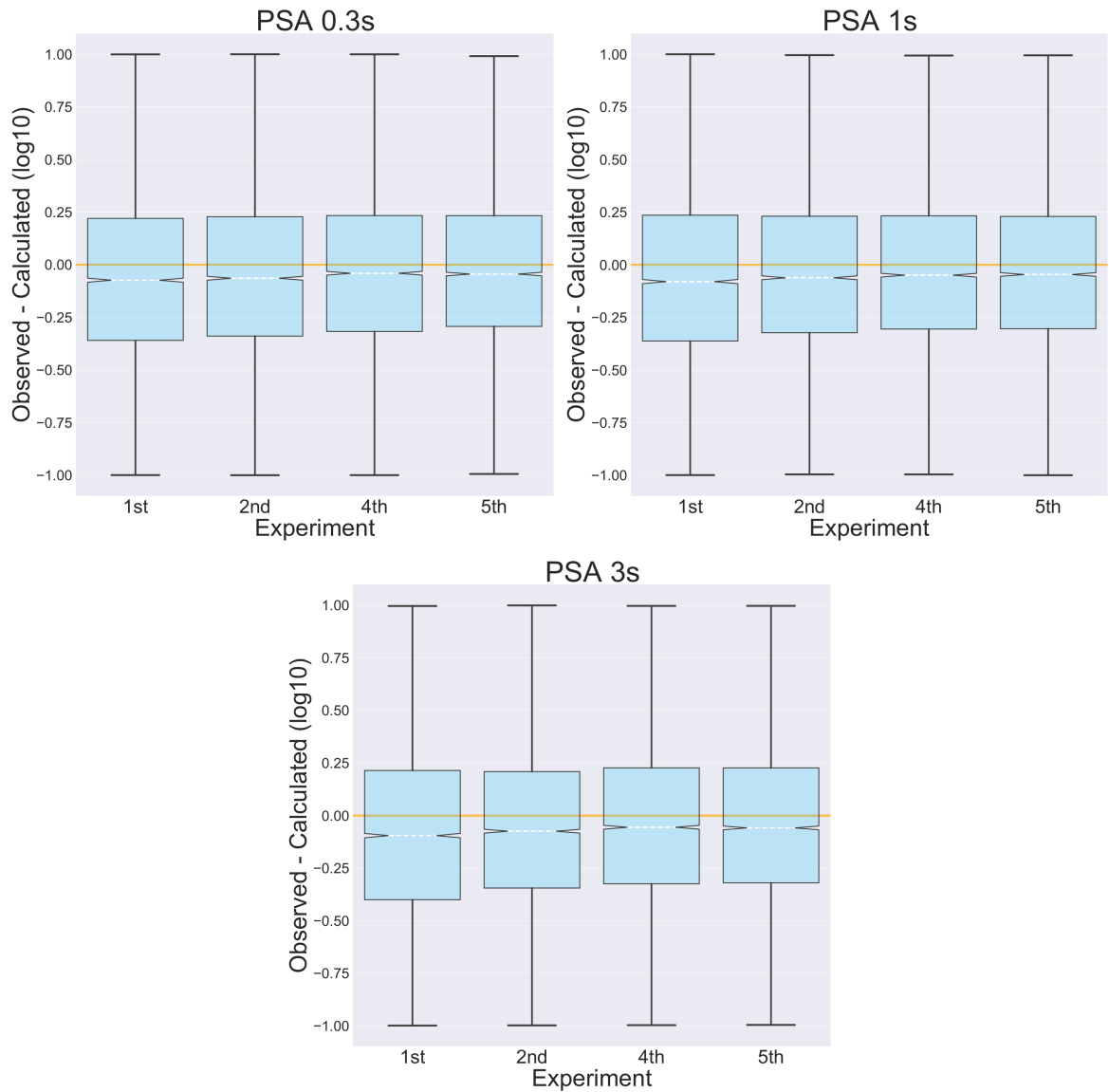

Figure 5. Boxplots for the residuals IM$_{observed}$ - IM$_{predicted}$ of the CNN training on the CW dataset from sections 4.1 (1st experiment), 4.2 (2nd experiment), 4.4 (4th experiment), and 4.5 (5th experiment) for different IMs: a) PGA b) PGV c)PSA 0.3 s, d) PSA 1.0 s, e) PSA 3.0 s

### 4.6. Results for station PII

We have extracted the residuals between the observed values and the CNN model predictions trained using TL described in section 4.5 for the station PII (part of the original set of 39 stations) since it is the closest station to the VIRGO detector site (coordinates:

43.63º N, 10.5º E) at about 10 km distance, which we use as an example of an infrastructure which would benefit from an EEWS. Using the same criteria for outliers as before, |R|>1, 89.1% of the data have been kept for observed IM targets and 91.8% for ShakeMap generated IM targets. The mean, median and standard deviations of the residuals are shown in table 6. We report the results for the observed targets and ShakeMap predicted targets separately and we note that the station PII had missing input waveforms for 42.9% of earthquakes. This is likely the reason why the results for the data with waveforms present show a deterioration of the results in the number of outliers and standard deviation compared to the data with input waveforms not present for the station PII. They also show slightly lower (absolute) bias, with the bias having opposite signs.

Table 6: IMs' residual R statistics for the 5th experiment: the CNN predictions on station PII from the CW dataset using TL from the pre-trained CI CNN model, with added interstation distances and azimuths.

| IM | Observed median | Observed mean | Observed STD | ShakeMap median | ShakeMap mean | ShakeMap STD |
|---|---|---|---|---|---|---|
| *PGA* | 0.146 | 0.181 | 0.392 | -0.23 | -0.202 | 0.355 |
| *PGV* | 0.099 | 0.097 | 0.351 | -0.164 | -0.180 | 0.321 |
| *SA(0.3)* | 0.219 | 0.214 | 0.356 | -0.283 | -0.230 | 0.367 |
| *SA(1.0)* | 0.042 | 0.07 | 0.368 | -0.141 | -0.083 | 0.34 |
| *SA(3.0)* | 0.008 | 0.022 | 0.352 | -0.097 | -0.036 | 0.38 |

## 4.7. Using station PII as the only model target

The previous results in section 4.6. are showing the results for station PII when the targets were the IMs at the 39 stations. In this section we report the results when we use only the station PII as the model target. We have calculated the residuals as in previous sections, and the same criterion as before was used to remove the outliers which led to keeping 92.76% for the observed IMs, and 93% for ShakeMap predictions. The results are reported in Table 7. Compared to section 4.6, where 39 stations were used as a target, we see an improvement in all the metrics for the observed target values. For the target labels derived from ShakeMap predictions, the results for the standard deviation are similar, whereas the bias and the number of outliers are reduced. The improvement in the data with the observed target labels led to results that were comparable to those with ShakeMap predicted target labels, contrary to the results of section 4.6, where the CNN model was performing better on the ShakeMap derived target labels.

Table 7: IMs' residual R statistics for the 6th experiment: the CNN predictions on station PII as the only target from the CW dataset using TL from the pre-trained CI CNN model, with added interstation distances and azimuths, for waveforms that end 7 seconds after the first P arrival at the closest station.

| IM | Observed median | Observed mean | Observed STD | ShakeMap median | ShakeMap mean | ShakeMap STD |
|---|---|---|---|---|---|---|
| *PGA* | 0.067 | 0.051 | 0.35 | -0.117 | -0.119 | 0.357 |
| *PGV* | 0.032 | 0.014 | 0.32 | -0.137 | -0.094 | 0.319 |
| *SA(0.3)* | 0.114 | 0.086 | 0.351 | -0.21 | -0.158 | 0.379 |
| *SA(1.0)* | 0.019 | -0.005 | 0.35 | -0.107 | -0.044 | 0.358 |
| *SA(3.0)* | -0.017 | -0.031 | 0.326 | -0.035 | -0.017 | 0.333 |

We also test how reducing the waveform window length affects the results by using the waveforms that end 4 seconds (i.e., total window length of 7 s) after the first P arrival at the closest station to the epicenter (in the previous sections the waveforms ended 7 seconds after the first P arrival at the closest station, i.e., 10 s total length). Using the same criteria for calculating the outliers we were left with 90.8% data kept for observed IMs and 89.5% data kept for ShakeMap predictions. The results are reported in table 8.

Table 8: IMs' residual R statistics for the 6th experiment: the CNN predictions on station PII as the only target from the CW dataset using TL from the pre-trained CI CNN model, with added interstation distances and azimuths, for waveforms that are 7 seconds long

| IM | Observed | Observed | Observed | ShakeMap | ShakeMap | ShakeMap |
|---|---|---|---|---|---|---|

|     | median | mean | STD | median | mean | STD |
|---|---|---|---|---|---|---|
| **PGA** | 0.09 | 0.071 | 0.386 | -0.105 | -0.102 | 0.366 |
| **PGV** | 0.046 | 0.032 | 0.367 | -0.072 | -0.066 | 0.335 |
| **SA(0.3)** | 0.133 | 0.121 | 0.378 | -0.178 | -0.128 | 0.376 |
| **SA(1.0)** | 0.009 | -0.007 | 0.379 | -0.04 | -0.038 | 0.370 |
| **SA(3.0)** | -0.033 | -0.029 | 0.35 | -0.025 | -0.01 | 0.378 |

We also calculate the predictions for the station PII using the GMPE of Bindi et al. (2011), for which we used the final earthquake location and magnitude. The predictions were calculated only for the earthquakes for which the station PII had recorded data. The residuals R were calculated as for the CNN predictions, and outliers were removed which led to keeping 85.3% of the data. The median, mean, and standard deviation of R are reported in table 9.

Table 9: IMs' residual R statistics for the GMPE predictions on station PII.

| IM | Observed median | Observed mean | Observed STD |
|---|---|---|---|
| **PGA** | 0.377 | 0.385 | 0.276 |
| **PGV** | 0.138 | 0.138 | 0.262 |
| **SA(0.3)** | 0.332 | 0.315 | 0.245 |
| **SA(1.0)** | 0.051 | 0.002 | 0.277 |
| **SA(3.0)** | 0.087 | 0.102 | 0.197 |

## 5. Discussion

In the study, we have shown that the introduction of TL in our CNN model can compensate for the lack of data and improve the results for rapid prediction of earthquake IMs., We have found that using a pre-trained CNN model, trained on CI in J2020, improved the results obtained using only the CW dataset (section 4.2, Fig. 5). Improvements (mean of the improvements of the 5 IMs) include: reducing the number of outliers by 4%, the residuals median by 19% and their standard deviation by 6%. These results suggest that TL is a viable technique to improve model performance on small-sized datasets. The convolutional filters in the first two layers of the CNN model (i.e., those used for TL) are the same for all the stations, which means that they have to be general enough to be able to extract features from the inputs regardless of which station they are currently analyzing. The third layer, which looks at cross station information (i.e., the station pattern of the ground motion and station-specific site amplifications due to soil type, topography, etc.), did not improve the results when included in TL for our problem. This was expected since the convolutional filters used in the third layer are of height 39 and span the specific geometrical/geographical pattern of the recording stations. Therefore, inserting a sequence of recording traces for TL unrelated to that target problem will worsen the results, unless the same geographical pattern is used and the stations reflect the same characteristics in terms of local site amplifications - a highly unlikely situation in practice. Given that our first two layers use single station filters, we could use the first two layers of the pre-trained model regardless of the number of stations used to create the input data. This can be useful if we want to train our model for an area that has a different number of stations available.

We also show in section 4.3 that parts of the CNN model, which was trained for a different problem (single station magnitude determination), can be used for TL on our CNN model, with an improvement on the results of the J2020 study in the number of outliers, the standard deviation of the residuals and their bias. Lower levels of improvement on the stations with missing data could be explained by the fact that the pre-trained models (i.e., trained on LEN-DB or STEAD) did not have stations in which the waveform data were not present, which is instead the case for 7% of the data in CI. It is, however, noteworthy that the CNN model was still able to compensate for the missing data even if the first two layers (pre-trained on STEAD and LEN-DB) were not pre-trained with missing data. Slightly better results using the STEAD pre-trained model, compared to the LEN-DB pre-trained model, could be explained by the higher sampling rate used with STEAD (100 Hz) compared to LEN-DB (20 Hz). The waveforms of LEN-DB, after the applied processing, were acceleration, i.e., the same units as the waveforms used for CI and CW, while STEAD waveforms were velocity waveforms in counts. However, as the first two layers of our model operate on normalized waveforms without absolute amplitude information, and the normalization constant is only inserted later in the model, the use of data in counts for pre-training the first two layers should not adversely affect the model performance.

Both LEN-DB and STEAD consist of earthquake waveforms recorded on available stations up to a specified distance from the earthquake epicenter, in contrast to our use of a fixed set of stations with waveforms for a set of earthquakes for training our IM prediction model. This difference essentially means that we cannot use these datasets to pre-train a CNN model that has the same, multi-station, architecture for IM prediction as we use in this study. However, this problem was circumvented by pre-training a single station and magnitude determination model instead, and then reusing its specific transferable layers for IM prediction, which is our target goal.

When using the pre-trained model trained on the same problem (CI IMs prediction, section 4.2), the best results were achieved when the learning rate of the first two layers was set to 0. When using the model trained on the different problem instead (magnitude determination, section 4.3), they were achieved when the first layer learning rate was set to 0 and the second remained trainable. This suggests that the features extracted by the first layer of the magnitude determination model can be directly used for IM prediction, without the need for further parameter fine-tuning. Model performance can be improved, however, by fine-tuning the second layer parameters to the target domain (i.e., the IM prediction). Moreover, further improvement achieved by fine-tuning all the layers showed that even the weights of the first layer, which was used from the pre-trained model, could be further adapted to better fit the new domain. This was achieved by using a smaller learning rate for fine-tuning, whereas, on the other hand, using greater learning rates did not improve model performance. When we tried to fine-tune the first two layers using a TL model in the same domain, the results did not improve, and fine-tuning led to overfitting.

The comparison of the results in sections 4.2 and 4.4 suggests that the weights of the first two pre-trained layers were already satisfactory for TL on our problem and that the training bottleneck was in the deeper layers of our model. The CNN model is learning the inter-station relations (locations, distances) and the characteristics of the stations implicitly during the training, as no station information has been given to it. Explicitly providing the inter-station distances and azimuths improved the results of the model (section 4.5). This can be understood as a form of TL, as we are guiding the model to obtain useful features that describe the data without needing to learn how to extract them. Normalizing the inter-station distances by the maximum of the distances improved the results, which follows the suggested normalization for neural network inputs (LeCun et al. 2012). The azimuth between the stations, *Az*, was initially provided in degrees and radians without any results improvement. The results were improved with the use of *sin(Az)* and *cos(Az)*, which are effectively normalizing azimuth description to the range [-1,1]. This azimuth description also gives the neural network a more meaningful measure of the closeness of two angles and removes the possible ambiguity that could come from using the angles (e.g., the angles 360° and 0° are the same angle, but the numerical difference between them is 360). Giving only the sine or cosine would also confuse the model, as they are both non-injective functions; however, providing both allows for an accurate description of the angle. In contrast, the addition of $v_{S,30}$ did not improve the results of the CNN model. Overall, when we compare the results in sections 4.5 and 4.1, we can see that the improvement from the use of TL comes in terms of reducing the number of outliers by 5%, the residuals R median by 41% and their standard deviation by 9%.

The test of our algorithm for EEW purposes was done on the station PII because it was the closest station (10 km) to the VIRGO gravitational wave observatory (Olivieri et al. 2019). To this end, we calculated the differences $T_p$ between the first P arrival times on the station PII (IV network) and the input waveforms start time, and show them in Fig. 6 (blue histogram bars). We can see in Fig. 6 that the CNN model would be able to give early warning for many earthquakes, depending on the length of the input window used. This rate is greater if we consider that the peak ground motions originate from S and surface waves, which arrive later, giving some more warning time before the strongest shaking. Therefore, in Fig. 6 we also show the differences $T_{pga}$ (light orange histogram bars) between the recorded PGA times at the station PII and the input waveform start times. These results do not account for the times needed for running the algorithm (very minor indeed) and for data transmission. It follows that the timing relation for $T_p$ and $T_{pga}$ shown in Fig. 6 support the use of the CNN model as an EEW system for VIRGO, as the large majority of $T_{pga}$ occur well after the 10 s input waveform end time (pink line in Fig. 6) so there would be warnings before the strongest shaking for a large majority of earthquakes. Moreover, if we reduce the input waveform length (section 4.7), we can see in Fig. 6 that these warnings for a large number of earthquakes could be given even before the first P arrival at PII (i.e., the large majority of $T_p$ are after the 7 s input waveform end time (green line)). An example earthquake input waveforms are shown in Fig. 1 of the electronic supplement, with an epicentral distance from station PII of 66 km, with $T_p$=9.42 s and $T_{pga}$=18.82 s.

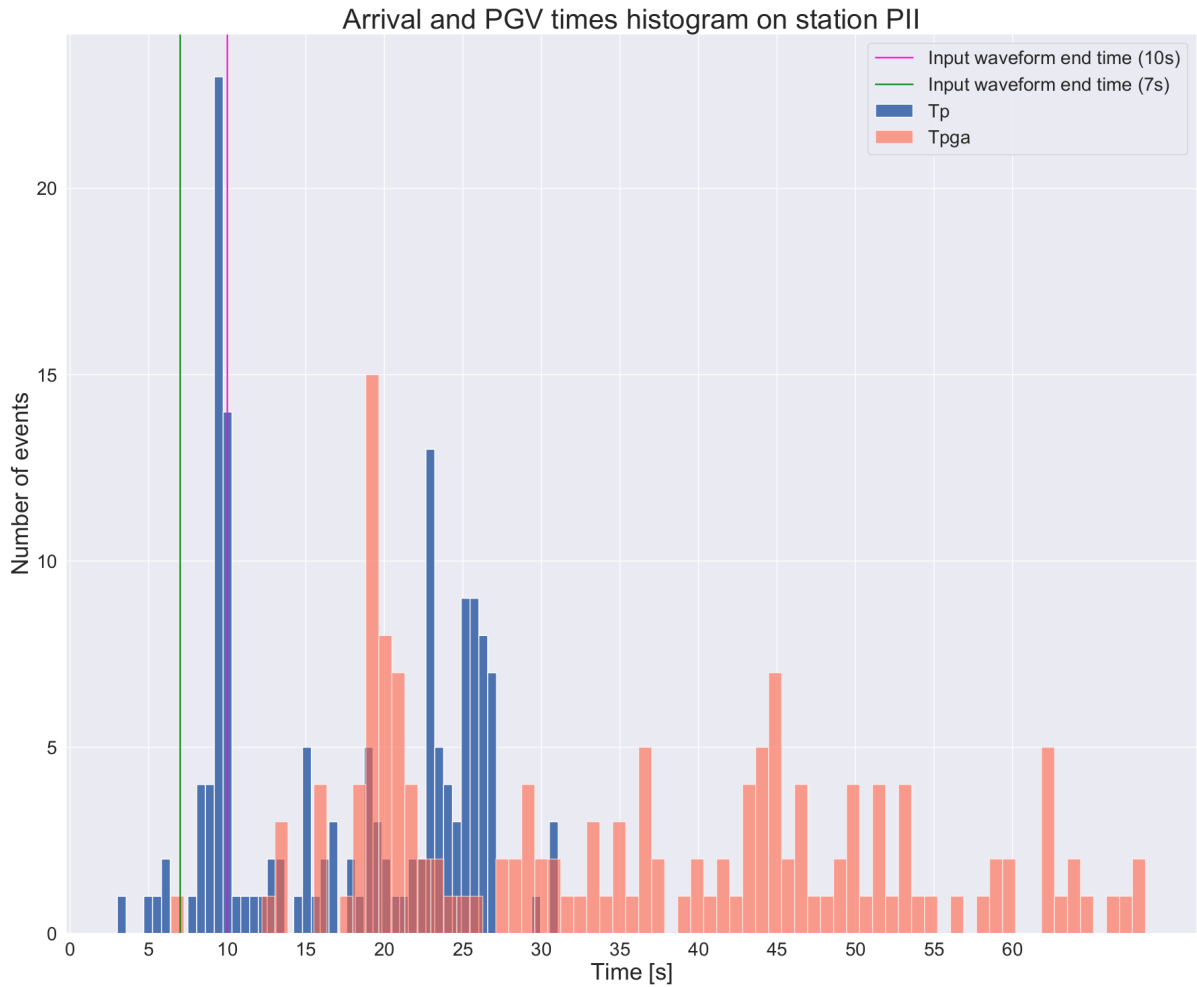

*Figure 6. Histograms of waveform arrival and peak ground motion delay times. Blue bars show the difference between the P arrival time at the station PII and the input waveform start time $T_p$. Light orange bars (reddish when overlapping with the $T_p$ blue bars) show differences between the recorded PGA time and the input waveform start time $T_{pga}$. Green and magenta vertical lines show the end of the 7 s and 10 s input waveforms, respectively. The bars on the right of those lines show the number of events, with possible warning times before the P arrival or the PGA time on station PII. The total number of events shown for $T_p$ is 266, while for the $T_{pga}$ only the 152 events with recorded waveforms (i.e. recorded PGA time) are shown.*

The results shown in section 4.6 suggest that the algorithm would give useful and timely predictions, for cases both with and without waveform data for station PII. Contrary to the results in the sections before, the CNN performs better on the data without the input waveforms present. The main difference in the case of station PII compared to the other stations used is the lower number of earthquakes for which the input waveforms are available (152 of 266 earthquakes). This means that it has less training data to learn how to predict the actual IMs recorded on the stations, and more data to learn how to predict from the ShakeMap derived values at those stations. It is also interesting to note that the CNN model is underpredicting the IMs for earthquakes for which the input waveforms exist, and overpredicting for earthquakes for which there are no input waveforms on the station PII, with the absolute values of mean and median smaller of the residuals R on the observed data. However, on the stations with missing data, the CNN results are being compared to ShakeMap predictions, which are calculated using a GMPE of Bindi et al. (2011) and geospatial interpolation of the observed values at the stations (Worden et al. 2020). We calculated the IMs on the station PII by using the GMPE of Bindi et al. (2011), which uses the final earthquake location and magnitude, for the earthquakes for which observed IMs exist. We then calculated the residuals R and found that the GMPE is also underpredicting the IM values, with a higher median than the CNN model on the same data. This suggests that the overprediction of CNN for the ShakeMap predictions comes from the CNN results having a smaller bias on recorded IM values.

When using only the station PII as a target, we can see that the results of the CNN are improved. This is expected because we are reducing the number of parameters (unknown variables) in the model. This suggests that for achieving the best results for a specific site of interest, individual training should be done. However, the training data does not need to be changed, only the output location. It is also interesting to note that the results were mostly improved, compared to section 4.6, for the earthquakes with recorded waveforms. Reducing the length of the window, expectedly, led to poorer model performance and larger uncertainties, although there is a trade-off between additional warning time achieved and loss in accuracy. An optimal ratio between accuracy and timeliness can be selected for every application individually. This trade-off also suggests that in a real-time application of the methodology progressively larger time windows could be employed after an earthquake occurs to obtain increasingly accurate predictions.

## 6. Conclusions

We used a CNN model to predict IMs (PGA; PGV; PSA 0.3 s, 1 s, 3 s) at 39 stations using multi-station waveforms that start 3 seconds before the first P arrival at the closest station and are 10 seconds long. The training set has only 266 earthquakes (with M > 3) and TL was used to improve the IMs prediction accuracy.

We show that TL is a powerful methodology to use in waveform data analysis for cases with insufficient training data. The TL can be done by using a pre-trained model trained on the same problem (IM prediction) or a different one (magnitude characterization) - with both cases improving the model accuracy. We also show that the inclusion of additional knowledge to the model (the interstation relations) can improve the training results.

We find with TL that the first two layers of the pre-trained model are the most important because they are used for feature extraction from single station inputs and that the learning rate of these layers can be set to 0. Therefore, when doing TL using our proposed model, only the third, cross-station layer needs to be retrained. This also implies that the adoption of

39 stations to construct the input waveforms for TL is arbitrary and can be varied according to the target problem, as we use TL only for the single-station feature extractors.

The experiment to predict the IMs at only one station (PII) showed that the simplification of the problem (i.e., reduced indeterminacy deriving from many fewer unknowns) led to performance improvements. We also show that in a possible application of our model for EEW at the VIRGO gravitational wave observatory, our CNN model could provide warnings for a large number of earthquakes of the selected dataset and a satisfying accuracy.

## 7. Data and resources

Earthquake catalogue and waveform data have been downloaded through the INGV FDSN web services (http://terremoti.ingv.it/en/webservices_and_software; INGV Seismological Data Centre 2006; Emersito Working Group 2018). Waveforms have been downloaded and processed using python library Obspy (Beyreuther et al. 2010). IMs for the stations with no data have been calculated using USGS ShakeMap 4 (http://usgs.github.io/shakemap/sm4_index.html). The CNN model has been developed using the Keras Python Deep Learning library (Chollet et al. 2015). The models and the data of this paper will be available on https://github.com/djozinovi/TLpredIM.

## 8. Acknowledgements

The research has been funded by SERA EU project (Seismology and Earthquake Engineering Research Infrastructure Alliance for Europe; contract n. 730900). It has also greatly benefited from a travel grant for a short-term scientific mission provided by G2NET,

COST action CA17137. The authors would like to thank the operators of RSN (INGV) and the USGS ShakeMap team of developers.## 8.1 Author contribution statement

# Electronic supplement

## 1. Additional Figures

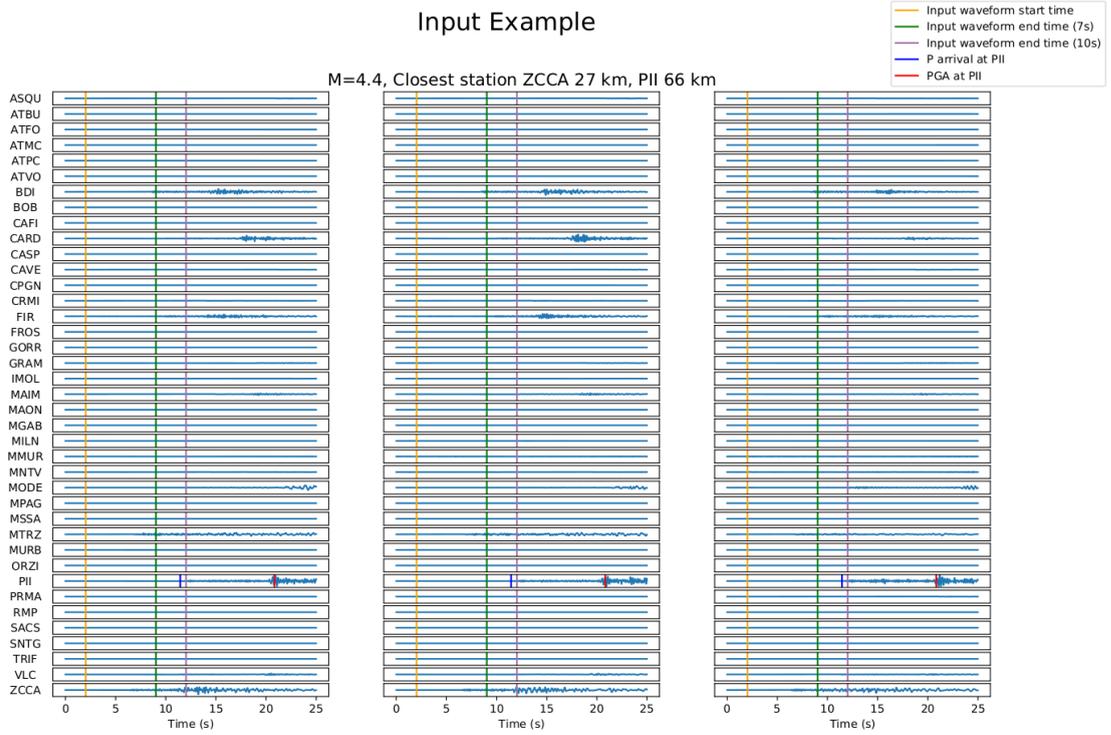

Figure 1. Input example with the normalized traces for a M=4.4 earthquake

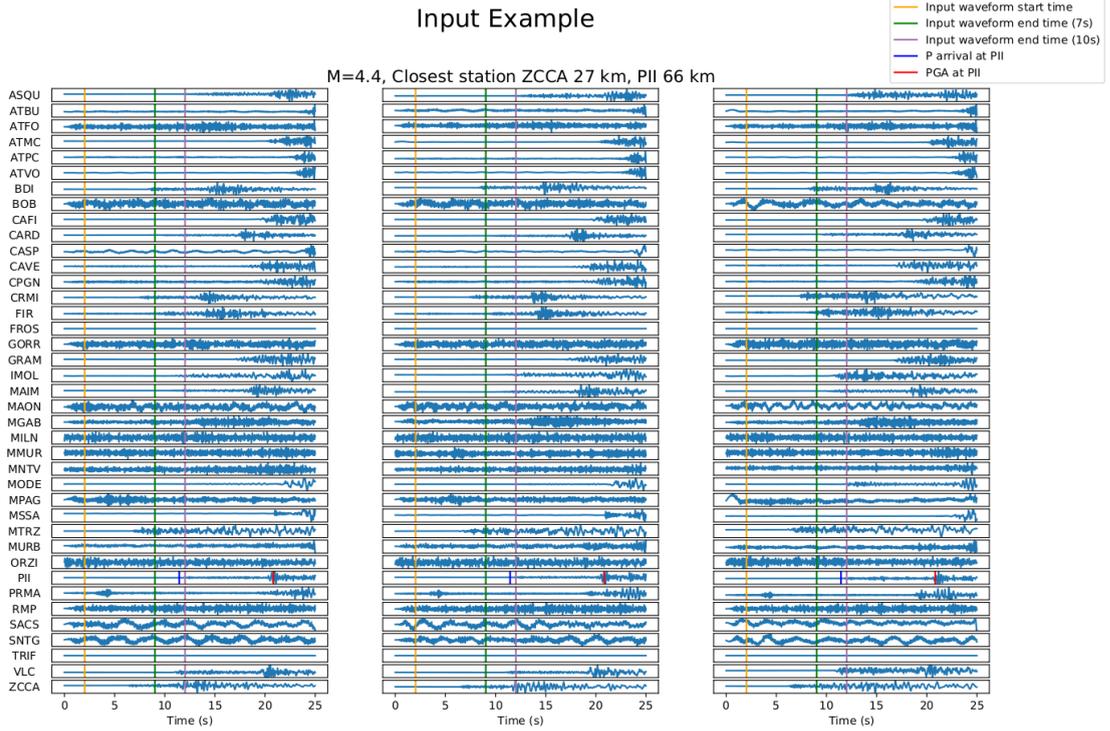

Figure 2. Non-normalised input example for a M=4.4 earthquake.

## 2. Station list

Table 1. The stations used in the study and some of their characteristics.

| Sta | Net | Lat | Lon | No. Eqs | PGA max | PGA min | PGA median |
|---|---|---|---|---|---|---|---|
| ASQU | IV | 43.7967 | 11.7893 | 253 | -0.783 | -5 | -3.686 |
| ATBU | IV | 43.4757 | 12.5483 | 247 | -0.208 | -5.057 | -3.744 |
| ATFO | IV | 43.3666 | 12.5715 | 259 | -0.534 | -5.028 | -3.847 |
| ATMC | IV | 43.4468 | 12.1928 | 251 | -1.217 | -5.021 | -3.688 |
| ATPC | IV | 43.4807 | 12.457 | 256 | -0.59 | -5.42 | -3.681 |
| ATVO | IV | 43.3821 | 12.4066 | 249 | -0.69 | -5.223 | -3.871 |
| BDI | IV | 44.0624 | 10.597 | 260 | 0.34 | -5.453 | -3.64 |
| BOB | IV | 44.7679 | 9.4478 | 249 | -1.727 | -9.598 | -3.975 |
| CAFI | IV | 43.3292 | 11.9663 | 255 | -2.045 | -9.598 | -3.827 |
| CARD | GU | 44.026 | 10.4821 | 253 | -0.457 | -4.92 | -3.407 |
| CASP | IV | 42.7908 | 10.8652 | 247 | -1.72 | -5.528 | -4.403 |
| CAVE | IV | 44.8658 | 11.0031 | 258 | -0.958 | -4.712 | -3.378 |
| CPGN | IV | 43.8011 | 12.3205 | 261 | -0.41 | -5.53 | -3.94 |
| CRMI | IV | 43.7956 | 10.9795 | 241 | -1.892 | -5.629 | -3.996 |
| FIR | IV | 43.7744 | 11.2551 | 238 | -0.886 | -4.537 | -2.241 |
| FROS | IV | 43.2097 | 11.1562 | 238 | -0.699 | -4.904 | -3.859 |
| GORR | GU | 44.6071 | 9.2926 | 256 | -0.554 | -5.71 | -4.063 |
| GRAM | GU | 44.4913 | 10.0658 | 258 | 1.233 | -4.68 | -3.332 |
| IMOL | IV | 44.3595 | 11.7425 | 254 | -1.944 | -5.325 | -4.556 |
| MAIM | GU | 43.9142 | 10.4915 | 263 | -1.085 | -5.318 | -3.814 |
| MAON | IV | 42.4283 | 11.1309 | 235 | -1.751 | -5.031 | -4.353 |
| MGAB | IV | 42.9126 | 12.1121 | 252 | -1.345 | -5.19 | -3.997 |
| MILN | IV | 45.4803 | 9.2321 | 252 | -0.59 | -5.62 | -3.023 |
| MMUR | IV | 43.4418 | 12.9973 | 247 | -1.173 | -4.842 | -2.907 |
| MNTV | IV | 45.1495 | 10.7897 | 251 | -1.483 | -4.28 | -2.83 |
| MODE | IV | 44.6297 | 10.9492 | 255 | -0.896 | -3.843 | -3.12 |
| MPAG | IV | 43.6292 | 12.7595 | 254 | -1.816 | -5.14 | -3.998 |
| MSSA | IV | 44.3162 | 9.5174 | 247 | -1.322 | -5.473 | -3.941 |
| MTRZ | IV | 44.3128 | 11.4248 | 253 | -0.943 | -4.593 | -3.736 |
| MURB | IV | 43.263 | 12.5246 | 253 | -0.59 | -4.964 | -3.509 |
| ORZI | IV | 45.4056 | 9.9307 | 255 | -1.82 | -5.091 | -3.464 |
| PII | IV | 43.7219 | 10.525 | 156 | -1.094 | -5.636 | -3.221 |
| PRMA | IV | 44.7637 | 10.3131 | 252 | -0.803 | -4.559 | -3.65 |
| RMP | IV | 41.8111 | 12.7022 | 252 | -2.858 | -4.896 | -3.887 |
| SACS | IV | 42.8491 | 11.9097 | 251 | -0.819 | -5.581 | -4.303 |
| SNTG | IV | 43.255 | 12.9406 | 248 | -1.464 | -5.166 | -4.126 |
| TRIF | IV | 43.1148 | 10.9026 | 227 | -1.165 | -9.598 | -4.076 |
| VLC | MN | 44.1594 | 10.3864 | 249 | -0.682 | -5.031 | -3.995 |
| ZCCA | IV | 44.3509 | 10.9765 | 252 | -1.454 | -5.36 | -3.966 |

## 3. Results separated into observed and not observed target labels

For each test, the CNN model results from all 5-fold cross-validation test sets are averaged. Residuals $R = \log_{10}(IM_{obs}/IM_{predicted})$ are calculated, then outlier residuals with $|R|>1$ are removed, and finally the mean, median and standard deviation of the outlier removed residuals are calculated. The results are separated into the residuals for CNN predictions on data that had observed target labels and those that were generated using ShakeMap.

### 3.1. Training from scratch

Removal of outlier residuals resulted in keeping 84.88% of the data for the observed IM targets, and 78.15% of the data for the ShakeMap generated targets.

Table 2: IMs' residual statistics for the CNN predictions for the observed IMs (for the stations having recorded data) and for the ShakeMap predictions (for the stations that had no recorded data).

| IM | Observed median | Observed mean | Observed STD | ShakeMap median | ShakeMap mean | ShakeMap STD |
|---|---|---|---|---|---|---|
| *PGA* | -0.081 | -0.065 | 0.412 | -0.121 | -0.124 | 0.421 |
| *PGV* | -0.106 | -0.088 | 0.407 | -0.078 | -0.081 | 0.415 |
| *SA(0.3)* | -0.071 | -0.055 | 0.410 | -0.107 | -0.092 | 0.421 |
| *SA(1.0)* | -0.084 | -0.06 | 0.412 | -0.048 | -0.051 | 0.416 |
| *SA(3.0)* | -0.1 | 0.083 | 0.420 | -0.032 | -0.035 | 0.447 |

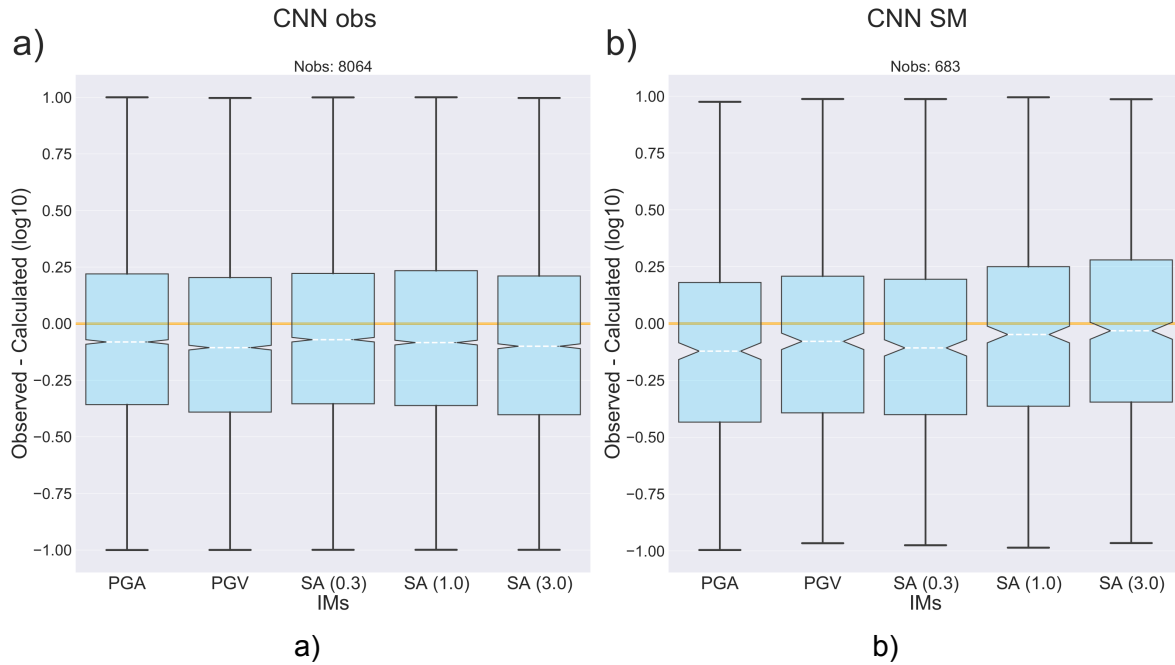

a)                                                          b)

Figure 3. Boxplots for the residuals $IM_{observed} - IM_{predicted}$: a) CNN model results for the observed IMs, b) CNN model results for the ShakeMap predicted IMs.

### 3.2. Transfer learning from a model pre-trained on central Italy dataset

The results of the CNN model trained using transfer learning from the pre-trained CNN central Italy model are presented here. Removal of outlier residuals resulted in keeping 88.55% of the data for the observed IM targets, and 81.58% of the data for the ShakeMap generated targets.

Table 3: IMs' residual statistics for the CNN predictions for the observed IMs (for the stations having recorded data) and the ShakeMap predictions (for the stations that had no recorded data).

| IM | Observed median | Observed mean | Observed STD | ShakeMap median | ShakeMap mean | ShakeMap STD |
|---|---|---|---|---|---|---|
| **PGA** | -0.068 | -0.052 | 0.394 | -0.076 | -0.077 | 0.399 |
| **PGV** | -0.087 | -0.071 | 0.378 | -0.05 | -0.046 | 0.396 |
| **SA(0.3)** | -0.063 | -0.044 | 0.4 | -0.07 | -0.047 | 0.413 |
| **SA(1.0)** | -0.065 | -0.041 | 0.393 | -0.018 | -0.014 | 0.419 |
| **SA(3.0)** | -0.08 | -0.066 | 0.388 | -0.003 | -0.0 | 0.422 |

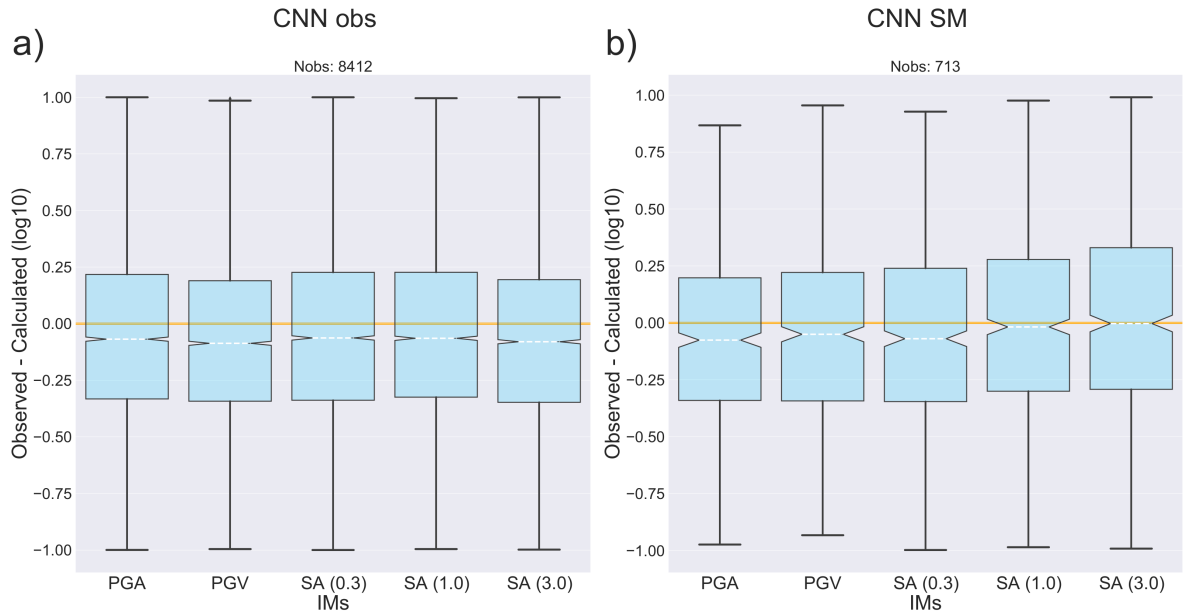

Figure 4. Boxplots for the residuals $IM_{observed} - IM_{predicted}$: a) CNN model results for the observed IMs, b) CNN model results for the ShakeMap predicted IMs

## 3.3. Improving the CIT model by transfer learning from a different task

Removal of the outliers resulted in 96.45% data kept for observed IM targets and 93.84% for ShakeMap generated IM targets.

Table 4: IMs' residual R statistics for the 3rd experiment: the CNN predictions on the CI dataset using TL from the pre-trained STEAD model. The results are reported for observed IMs (for the stations having recorded data) and the ShakeMap predictions (for the stations that had no recorded data).

| IM | Observed median | Observed mean | Observed STD | ShakeMap median | ShakeMap mean | ShakeMap STD |
|---|---|---|---|---|---|---|
| **PGA** | -0.011 | -0.005 | 0.276 | -0.010 | -0.025 | 0.331 |
| **PGV** | -0.017 | -0.005 | 0.252 | -0.046 | -0.040 | 0.317 |
| **SA(0.3)** | -0.005 | -0.0003 | 0.28 | -0.011 | -0.011 | 0.334 |
| **SA(1.0)** | -0.007 | 0 | 0.266 | -0.058 | -0.041 | 0.319 |
| **SA(3.0)** | -0.013 | 0 | 0.287 | -0.091 | -0.075 | 0.352 |

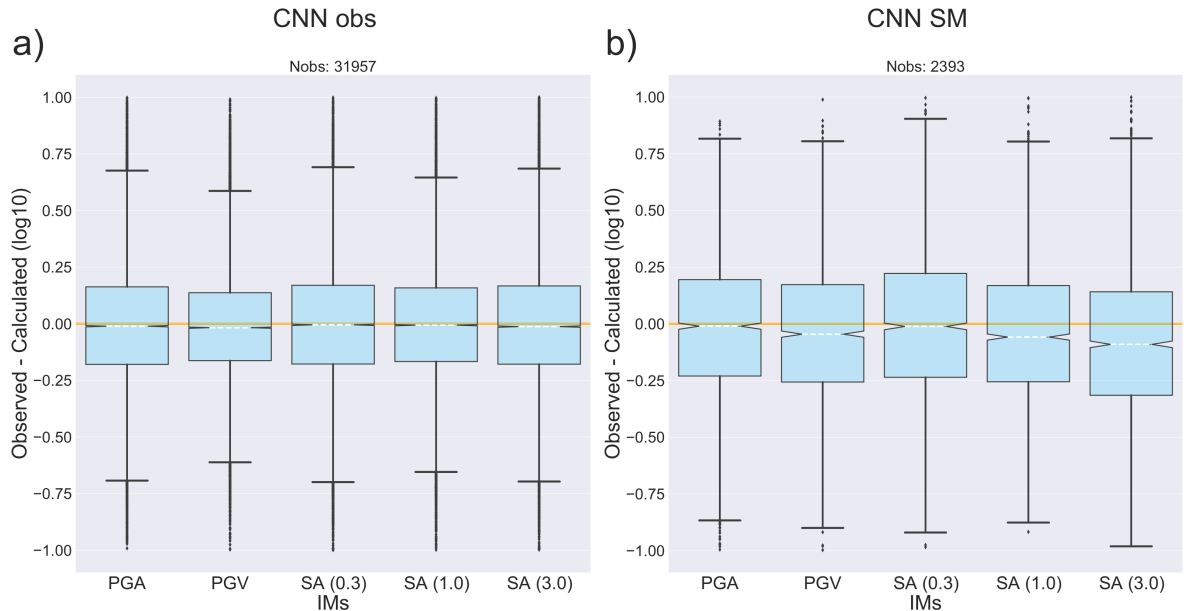

Figure 5. Boxplots for the residuals $IM_{observed}$ - $IM_{predicted}$ of the CNN model trained on the CI data using TL from a model pre-trained for magnitude prediction on STEAD: a) CNN model results for the observed IMs, b) CNN model results for the ShakeMap predicted IMs

We also report the results from the study of Jozinović et al. (2020) trained on CIT data in Table 5. Large residual values |R|>1 were removed resulting in 87.49 per cent of the data kept for the ShakeMap predictions and 92.00 per cent for the observed IMs.

Table 5: IMs' residual statistics for the CNN predictions for the observed IMs (for the stations having recorded data), ShakeMap predictions (for the stations that had no recorded data) and the predictions of GMPE by Bindi et al. from the study Jozinović et al. (2020)

| IM | Observed median | Observed mean | Observed STD | ShakeMap median | ShakeMap mean | ShakeMap STD | GMPE median | GMPE mean | GMPE STD |
|---|---|---|---|---|---|---|---|---|---|
| **PGA** | 0.038 | 0.035 | 0.346 | 0.059 | 0.038 | 0.372 | 0.013 | 0.017 | 0.352 |
| **PGV** | 0.036 | 0.034 | 0.338 | 0.043 | 0.041 | 0.380 | -0.174 | -0.151 | 0.33 |
| **SA(0.3)** | 0.031 | 0.031 | 0.34 | 0.056 | 0.046 | 0.37 | -0.284 | -0.252 | 0.359 |
| **SA(1.0)** | 0.029 | 0.034 | 0.338 | 0.001 | 0.017 | 0.374 | -0.207 | -0.198 | 0.303 |
| **SA(3.0)** | 0.019 | 0.027 | 0.374 | -0.037 | -0.012 | 0.404 | 0.026 | 0.083 | 0.368 |

## 3.4. Using the newly trained Central Italy model for transfer learning

When the outliers have been removed 87.84% data has been kept for the observed IM targets and 83.07% for SM generated IM targets.

Table 6: IMs' residual statistics for the CNN predictions for the observed IMs (for the stations having recorded data) and the ShakeMap predictions (for the stations that had no recorded data).

| IM | Observed median | Observed mean | Observed STD | ShakeMap median | ShakeMap mean | ShakeMap STD |
|---|---|---|---|---|---|---|
| *PGA* | -0.053 | -0.045 | 0.393 | -0.074 | -0.063 | 0.401 |
| *PGV* | -0.068 | -0.058 | 0.379 | -0.032 | -0.031 | 0.398 |
| *SA(0.3)* | -0.039 | -0.034 | 0.397 | -0.06 | -0.03 | 0.420 |
| *SA(1.0)* | -0.052 | -0.037 | 0.388 | -0.017 | 0.01 | 0.41 |
| *SA(3.0)* | -0.058 | -0.049 | 0.383 | -0.004 | -0.002 | 0.413 |

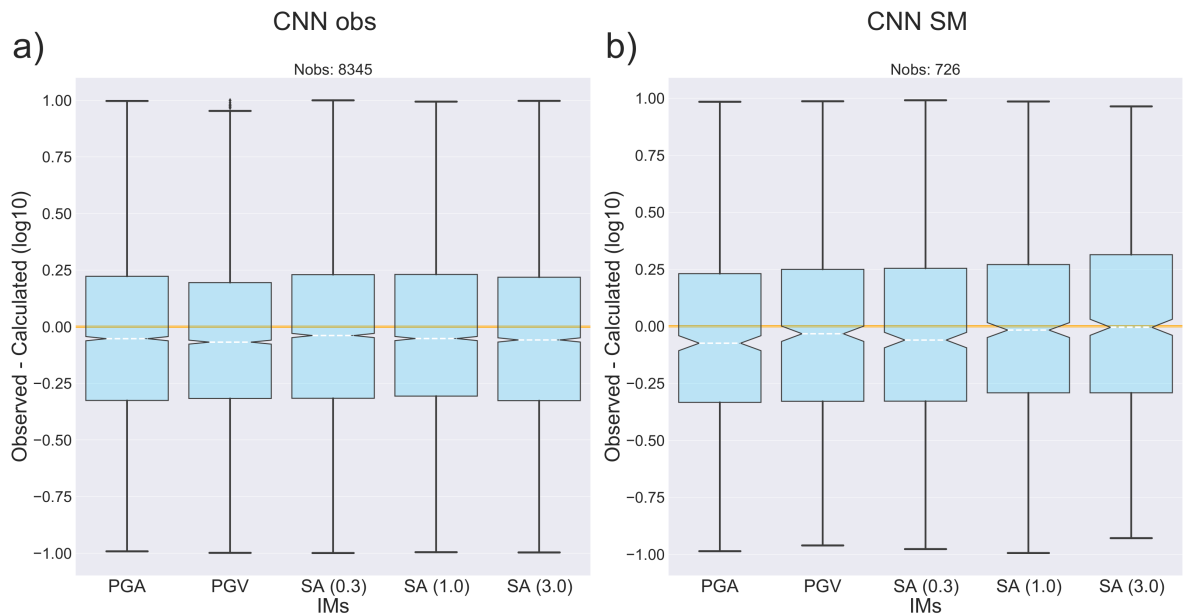

Figure 6. Boxplots for the residuals $IM_{observed} - IM_{predicted}$: a) CNN model results for the observed IMs, b) CNN model results for the ShakeMap predicted IMs

## 3.5. Adding additional metadata

When the outliers have been removed 89.83% data has been kept for the observed IM targets and 85.24% for SM generated IM targets.

Table 7: IMs' residual statistics for the CNN predictions for the observed IMs (for the stations having recorded data) and the ShakeMap predictions (for the stations that had no recorded data).

| IM | Observed median | Observed mean | Observed STD | ShakeMap median | ShakeMap mean | ShakeMap STD |
|---|---|---|---|---|---|---|
| *PGA* | -0.049 | -0.034 | 0.373 | -0.062 | -0.052 | 0.39 |
| *PGV* | -0.063 | -0.053 | 0.363 | -0.024 | -0.024 | 0.382 |
| *SA(0.3)* | -0.045 | -0.025 | 0.38 | -0.034 | -0.019 | 0.397 |
| *SA(1.0)* | -0.050 | -0.032 | 0.384 | 0.019 | 0.003 | 0.397 |
| *SA(3.0)* | -0.064 | -0.047 | 0.383 | 0.006 | 0.009 | 0.411 |

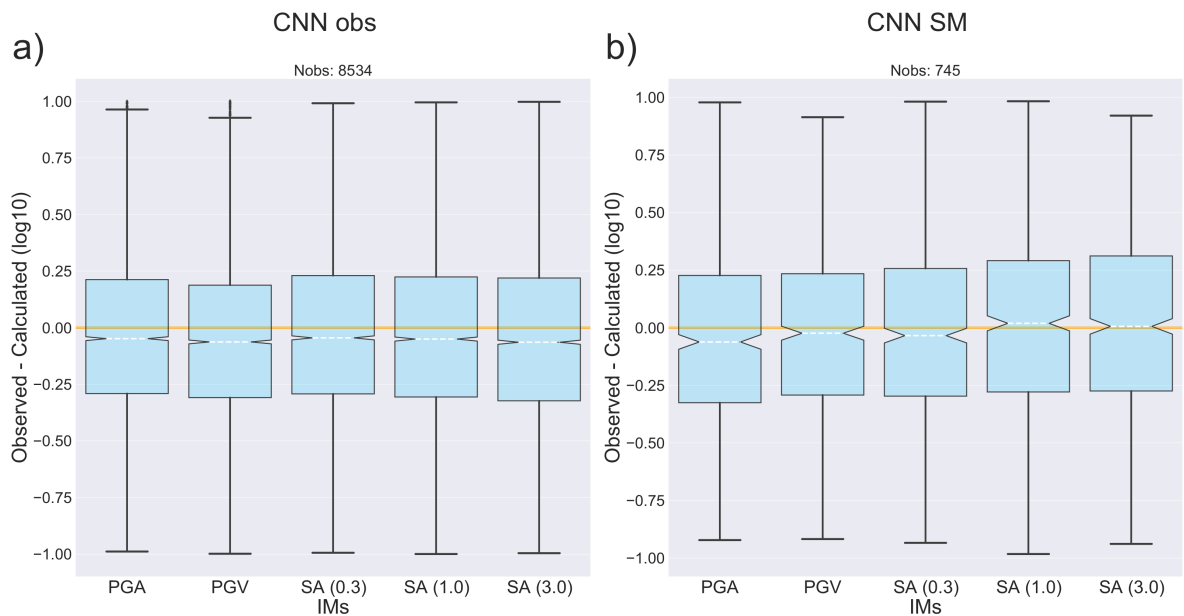

Figure 7. Boxplots for the residuals $IM_{observed} - IM_{predicted}$: a) CNN model results for the observed IMs, b) CNN model results for the ShakeMap predicted IMs